\RequirePackage{fix-cm}
\documentclass[smallextended]{svjour3}       
\smartqed  
\usepackage{graphicx}
\begin{document}

\title{Anthropic considerations in nuclear physics\thanks{
Work supported in part by DFG
and NSFC (Sino-German CRC 110), Helmholtz Association 
(contract VH-VI-417), BMBF\ (grant 05P12PDFTE),  the EU (HadronPhysics3 project),
and by LENPIC (DEC-2103/10/M/ST2/00420).
Computational resources provided 
by the J\"{u}lich Supercomputing Centre (JSC) at the Forschungszentrum 
J\"{u}lich  and by RWTH Aachen.}
}


\author{ Ulf-G. Mei{\ss}ner}


\institute{Ulf-G. Mei{\ss}ner\at
              Helmholtz-Institut f\"ur Strahlen- und Kernphysik (Theorie) and Bethe Center for Theoretical Physics, 
              Universit\"at Bonn, D-53155 Bonn, Germany\\
              \email{meissner@hiskp.uni-bonn.de}  
               \at Institute for Advanced Simulation (IAS-4), Institut f\"ur Kernphysik (IKP-3), J\"ulich Center for
                Hadron Physics, 
              JARA HPC and JARA FAME, Forschungszentrum J\"ulich, D-52425 J\"ulich, Germany,
              \at Kavli Institue for Theoretical Physics China, CAS, Beijing, 100190, China      
          }

\date{Received: date / Accepted: date}

\maketitle

\begin{abstract}
In this short review, I discuss the sensitivity of the generation of the light and the life-relevant elements like carbon and oxygen
under changes of the parameters of the Standard Model pertinent to nuclear physics. Chiral effective field theory allows
for a systematic and precise description of the forces between two, three, and four nucleons. In this framework, variations under the
light quark masses and the electromagnetic fine structure constant can also be consistently calculated. Combining chiral nuclear effective
field theory with Monte Carlo simulations allows to further  calculate the properties of nuclei, in particular of the Hoyle state in carbon,
that plays a crucial role in the generation of the life-relevant elements in hot, old stars. The dependence of the triple-alpha process on the fundamental constants of Nature is calculated  and some implications for our anthropic view of the Universe are discussed. 
\end{abstract}

\section{A brief guide through this short review}
\label{sec:intro}

In this review, I discuss certain fine-tunings in nuclear physics that are relevant to the formation of life-relevant
elements in the Big Bang and in stars. To set the stage, in Sec.~\ref{sec:antro} I give a brief discussion of the
so-called anthropic principle and argue that one can indeed perform physics tests of this rather abstract statement
for specific processes like  element generation. This can be done with the help of high performance computers that allow
us to simulate worlds in which the fundamental parameters underlying nuclear physics take values different from the
ones in Nature. In Sec.~\ref{sec:def} I define the specific physics problems we want to address, namely how sensitive the
generation of the light elements in the Big Bang is to changes in the light quark mass 
$m_q$\footnote{Throughout this review, we work in two-flavor QCD with up and down quarks with  masses $m_u$ and 
$m_d$, respectively.  In most cases, it suffices to 
work in the isospin limit $m_u = m_d \equiv m_q$ but at one instance we also have to consider strong isospin breaking 
with $m_u \neq m_d$.} and also, how robust the
resonance condition in the triple alpha process, i.e. the closeness of the so-called Hoyle state to the energy of $^4$He+$^8$Be, 
is under variations in  $m_q$ and the electromagnetic fine structure constant $\alpha_{\rm EM}$. The theoretical framework to
perform such calculations is laid out in Secs.~\ref{sec:nucl}~and~\ref{sec:nls}. First, I briefly discuss how the forces between nucleons
can be systematically and accurately  derived from the chiral Lagrangian of QCD. Second, I show how combining these
forces with computational methods allows for truly {\em ab initio} calculations of nuclei. In this framework, the decades old
problem of computing the so-called Hoyle state, a particular resonance in the spectrum of the $^{12}$C nucleus,
 and its properties can be solved. This is a necessary ingredient to tackle the problem
of the fine-tuning mentioned before. In Sec.~\ref{sec:forces}, I show how the quark mass dependence of the nuclear forces 
can be consistently calculated within chiral nuclear effective field theory (EFT). Constraints on such variations can be derived
from Big Bang nucleosynthesis, as outlined in Sec.~\ref{sec:bbn}. Here, we will encounter the first fine-tuning relevant to 
life on Earth. This, however, requires also heavier elements like carbon and oxygen. The viability of the generation of these
elements under changes in the light quark mass and the fine structure constant is discussed in Sec.~\ref{sec:fate}. I summarize
the implications of these findings for the anthropic principle in Sec.~\ref{sec:ant} and give a short summary and outlook
in Sec.~\ref{sec:sum}. I note that much more work has been done on the topics discussed here, for recent works and
reviews the reader is referred to Refs.~\cite{Jaffe:2008gd,Barnes:2011zh,Schellekens:2013bpa} and the papers quoted therein.

\section{The anthropic principle}
\label{sec:antro}

\begin{figure}[t]
\begin{center}
\includegraphics[width=0.40\textwidth]{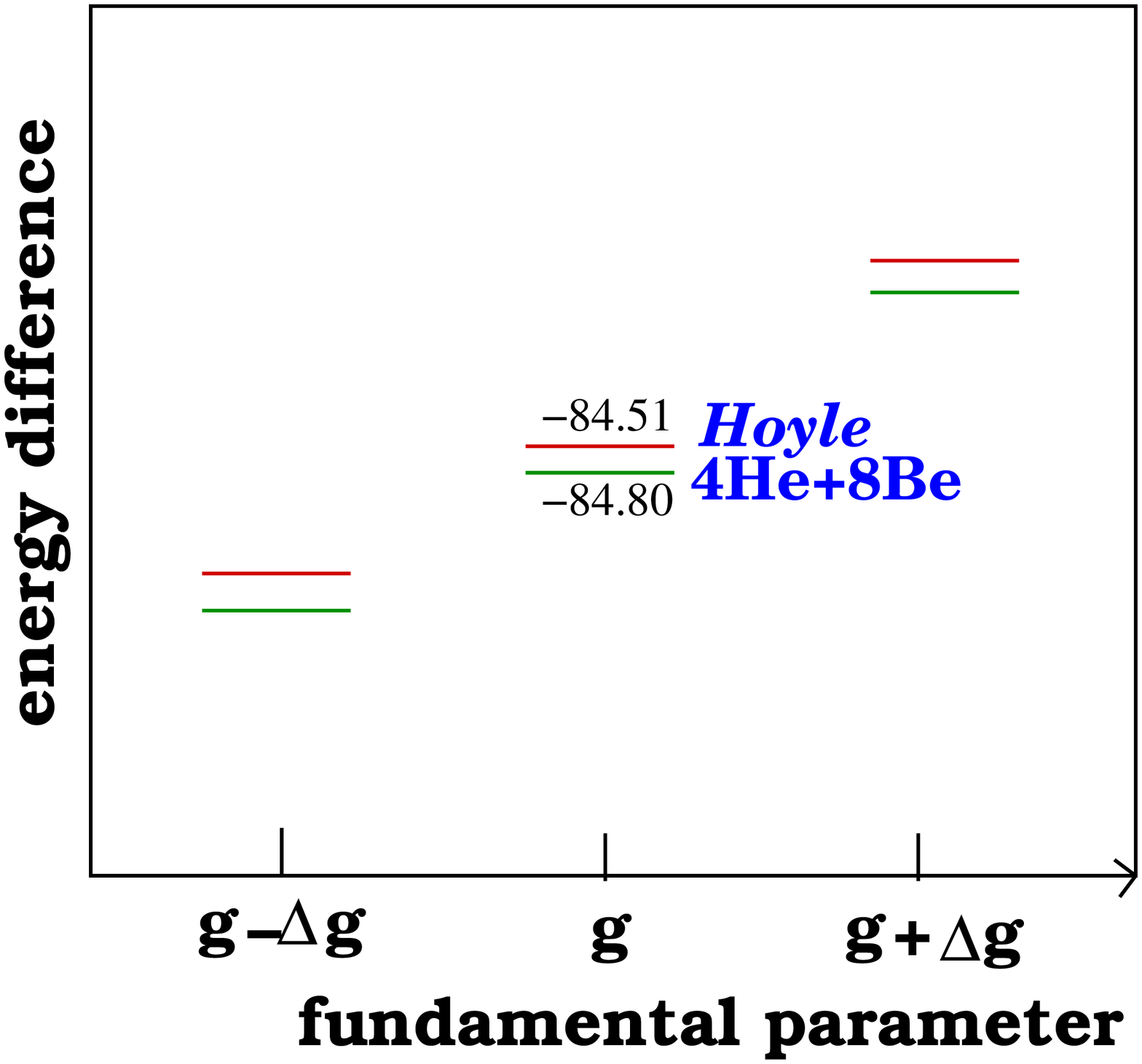}~~~~~~~`
 \includegraphics[width=0.40\textwidth]{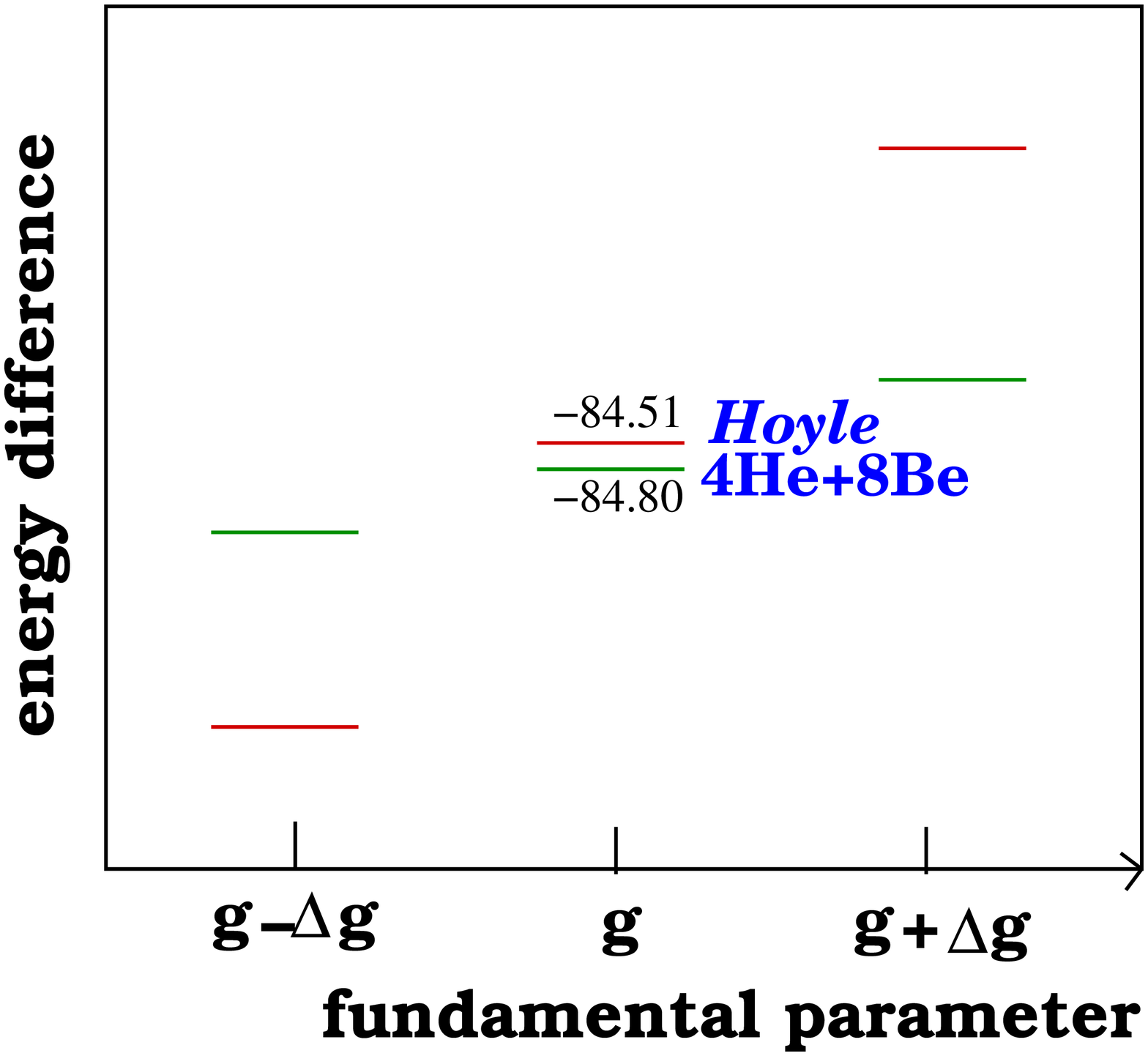}
\end{center}
\caption{Resonance condition for carbon production (closeness of the Hoyle state to the $^4$He+$^8$Be
threshold) in stars as a function of some fundamental
parameter $g$. Left panel: Non-anthropic scenario, right panel: anthropic scenario.}
\label{fig:para}
\end{figure}

The Universe we live in is characterized by certain parameters that take specific values
so that life on Earth is possible. For example, the age of the Universe must be large enough
to allow for the formation of galaxies, stars and planets. On more microscopic scales, certain
fundamental parameters of the Standard Model of the strong and electroweak interactions
like the light quark masses or the electromagnetic fine structure constant must take values that
allow for the formation of neutrons, protons and atomic nuclei. At present, we do not have a
viable theory to predict the precise values  of these constants, although string theory promises
to do so in some distant future. Clearly, one can think of many universes, the multiverse, in which
various fundamental parameters take different values leading to environments very different from ours.
In that sense, our Universe has a preferred status, and this was the basis of the so-called {\it anthropic
principle} (AP) invented by Carter \cite{Carter}. The AP states that ``the observed values of all physical 
and cosmological quantities are not equally probable but they take on values 
restricted by the requirement that there exist sites where carbon-based life can evolve and by the requirements that 
the Universe be old enough for it to have already done so''. There are many variants of the AP, but this
definition serves our purpose quite well. At first sight, one might think that it is a triviality, as the statement seems
to be a tautology. However, we can move away from the philosophical  level and ask whether the AP
can have physical consequences that can be tested? This is indeed the case particularly in nuclear
physics, as I will argue in this review. But it is worth mentioning that anthropic reasoning has been
used in some well cited papers, I name here Weinberg's work on the cosmological constant \cite{Weinberg:1987dv}
and Susskind's exploration of the string theory landscape \cite{Susskind:2003kw}.  
The influence of the AP on string theory and particle physics has been  reviewed recently in 
Ref.~\cite{Schellekens:2013bpa}. But let us return to nuclear physics. A prime
example of the AP is the so-called Hoyle state. In 1954, Hoyle \cite{Hoyle}
made the prediction of an excited 
level in carbon-12 to allow for a sufficient production of heavy elements  ($^{12}$C, $^{16}$O,...) in stars.
As the Hoyle state is crucial to the formation of the elements essential to life as we know it, this state
has been nicknamed the ``level of life''~\cite{Linde}.  See, however, Ref.~\cite{Kragh} for a thorough historical
discussion of the Hoyle state in view of the anthropic principle. Independent of these historical issues, 
the anthropic view of the Universe can be nicely shown using
the example of the Hoyle state, more precisely, one can understand how the abstract principle can be
turned into a physics question. The central issue is the closeness of the Hoyle state to the threshold
of $^4$He+$^8$Be that determines the resonance enhancement of carbon production. In Fig.~\ref{fig:para}
I show the possible response of this resonance condition to the change of some fundamental parameter,
here called $g$. If for a wide range of this parameter, the resonance condition stays intact (left panel), more precisely,
the absolute energies might shift but the Hoyle state stays close to the energy of $^4$He+$^8$Be. In such a case,
one can hardly speak of an anthropic selection. If on the other hand, the two levels split markedly for small
changes in $g$ as shown in the right panel, this would correspond to a truly anthropic fine-tuning. 
In Nature, we can not investigate which of these scenarios is indeed
fulfilled as all fundamental constants take specific values. However, with the powerful tool of 
computer simulations this has become possible and this issue will 
be  discussed in the remaining part of the review.

\section{Definition of the physics problem}
\label{sec:def}

In this section, I will more precisely define the nuclear physics problems
that have implications for our anthropic or non-anthropic view of the Universe.
As it is well known, the elements that are pertinent to life on Earth are generated in the 
Big Bang and in stars through the fusion of protons, neutrons and nuclei.
In Big Bang nucleosynthesis (BBN), alpha particles ($^4$He nuclei) and some
other light elements are generated. Life essential elements like $^{12}$C and
$^{16}$O are generated in hot, old stars, where the so-called triple-alpha
reaction plays an important role. Here, two alphas fuse to produce the
unstable, but long-lived $^8$Be nucleus. As the density of $^4$He nuclei in
such stars is high, a third alpha fuses with this nucleus before it decays.
However, to generate a sufficient amount of $^{12}$C, an excited
state in $^{12}$C at an excitation energy of 7.65~MeV with spin zero and
positive parity is required \cite{Hoyle}, this is the famous Hoyle state (for a 
recent review on the Hoyle state, see Ref.~\cite{Freer:2014qoa}).
In a further step, carbon is turned into oxygen without such a resonant
condition. So we are faced with a multitude of fine-tunings which need to
be explained. We know that  all strongly interacting composites like 
hadrons and nuclei must emerge from the underlying gauge theory of the
strong interactions, Quantum
Chromodynamics (QCD), that is formulated in terms of quarks and gluons.
These fundamental matter and force fields are, however, confined. Note that
the mass of the light quarks relevant for nuclear physics is very small ($m_u \simeq 2$~MeV
and $m_d \simeq 4$~MeV in the $\overline{\rm MS}$ scheme at $\mu = 2$~GeV) 
and thus plays little role in the total mass of nucleons and nuclei. However, the
light quark masses are of the same size as the binding energy per nucleon. 
Further, the formation of nuclei from neutrons and protons requires the inclusion of
electromagnetism, characterized by the fine-structure constant $\alpha_{\rm
  EM}\simeq 1/137$. So the question we want to address in the following is:
How sensitive are these strongly interacting composites to variations in the
fundamental parameters of the Standard Model? or stated differently: how accidental is
life on Earth?

\section{Chiral symmetry and nuclear forces }
\label{sec:nucl}

\begin{figure}[t]
\begin{center}
\includegraphics[width=0.9\textwidth]{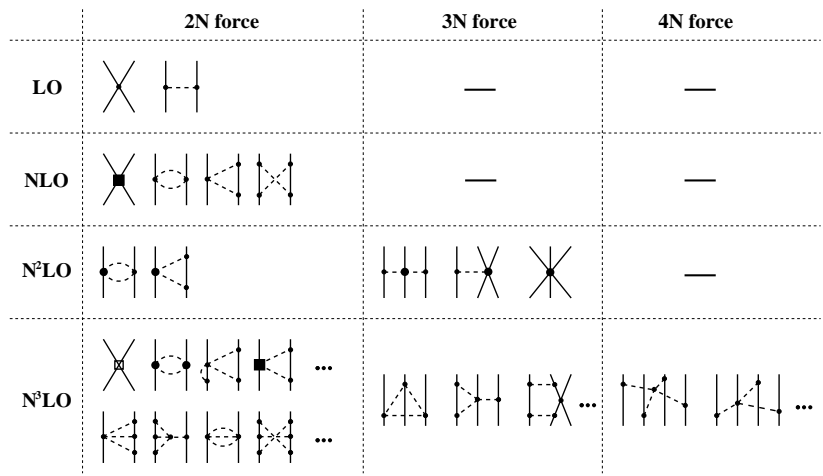}
\end{center}
\caption{\label{fig:power}
Contributions to the effective potential of the 2N, 3N and 4N forces
based on Weinberg's power counting.
Here, LO denotes leading order, NLO next-to-leading order and so on.
Dimension one, two and three pion-nucleon interactions are denoted
by small circles, big circles and filled boxes, respectively. In the
4N contact terms, the filled and open box denote two- and four-derivative
operators, respectively.
}
\end{figure}

It is known since long that  chiral symmetry  plays an important
role in a consistent and precise description of the forces between
nucleons. However,  a truly systematic approach based on the chiral effective 
Lagrangian of QCD only became available through the groundbreaking
work of  Weinberg \cite{Weinberg:1990rz}. As realized by Weinberg,
the power counting of the underlying EFT does not apply  directly to the
S-matrix, but rather to the effective potential - these are all diagrams
without $N$-nucleon intermediate states. Such diagrams lead to pinch 
singularities in the infinite nucleon mass limit (the so-called static limit), 
so that e.g. the  nucleon box graph is enhanced as $m_N/Q^2$, with 
$m_N$ the nucleon mass
and $Q$ a small momentum. The  power counting formula for the
graphs contributing with the $\nu^{\rm th}$ power of $Q$ or a pion mass  
to the effective potential reads (considering only connected pieces):
\begin{equation}
\nu = -2 + 2N + 2L + \sum_i V_i \Delta_i ~, \quad \Delta_i\ =  d_i + \frac{n_i}{2} -2 ~.
\end{equation}
Here, $N$ is the number of in-coming and out-going nucleons, $L$ the
number of pion loops, $V_i$ counts the vertices of type $i$ with $d_i$
derivatives and/or pion mass insertions and $n_i$ is the number of nucleons
participating in this kind of vertex. Because of chiral symmetry, $\Delta_i \geq 0$, 
and thus the  leading terms contributing e.g. to the two-nucleon potential can easily be 
identified. These are the time-honored one-pion exchange and
two four-nucleon contact interactions without derivatives. 
They can be derived from the lowest order effective chiral Lagrangian
with $\Delta_i = 0$ as indicated by the superscript '$(0)$',
\begin{eqnarray}
\label{lagr}
 \mathcal{L}^{(0)} &=& \frac{1}{2} \partial_\mu \vec\pi \cdot \partial^\mu \vec\pi - 
\frac{1}{2}  M_\pi^2 \vec\pi^2 + N^\dagger \left[ i \partial_0 + \frac{g_A}{2 F_\pi} \vec \tau \vec \sigma 
\cdot \vec \nabla \vec \pi - \frac{1}{4 F_\pi^2} \vec\tau \cdot (\vec \pi \times \dot{\vec \pi}) \right] N\nonumber\\
&& {} - \frac{1}{2} C_S (N^\dagger N) (N^\dagger N) -  \frac{1}{2} C_T (N^\dagger \vec \sigma N) \cdot (N^\dagger 
\vec \sigma N) + \ldots \,, 
\end{eqnarray}
where $\vec \pi$ and $N$ refer to the pion and nucleon field operators, respectively, 
and $\vec \sigma$ ($\vec \tau$) denote the spin (isospin) Pauli matrices.  Further, $g_A$ ($F_\pi$) is the nucleon 
axial coupling (pion decay) constant and $C_{S,T}$ are the LECs
accompanying the leading contact  operators without derivatives. 
The ellipses refer to terms involving more pion fields. It is
important to emphasize that chiral symmetry leads to highly nontrivial 
relations between the various coupling constants. For example, the strengths of all $\Delta_i=0$-vertices without 
nucleons with $2, 4, 6, \ldots$ pion field operators are given in terms of $F_\pi$ and $M_\pi$. Similarly, 
all single-nucleon $\Delta_i=0$-vertices with $1,2,3,\ldots$ pion fields are expressed in terms of just 
two LECs, namely $g_A$ and  $F_\pi$.   The corrections to the potential are then generated from the
higher order terms in the Lagrangian. The so-constructed effective potential is iterated
in the Schr\"odinger or Lippman-Schwinger equation, generating
the shallow nuclear bound states as well as scattering states. This requires regularization, a topic still under
current debate, but I do not want to enter this issue here, see e.g. Ref.~\cite{Epelbaum:2008ga}. 

\begin{figure}[t]
\begin{center}
 \includegraphics[width=0.95\textwidth]{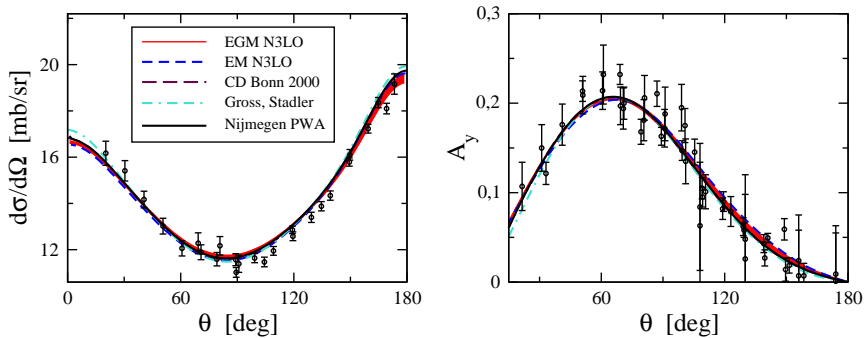}
\end{center}
\caption{Neutron-proton differential cross section $d\sigma/d\Omega$ (left panel) and
         analyzing power $A_y$ (right panel) at $E_{\rm lab} = 50$~MeV
         calculated using chiral EFT  at N$^3$LO by Epelbaum, Gl\"ockle and Mei{\ss}ner (EGM)
          \cite{Epelbaum:2004fk}   and Entem and Machleidt (EM) \cite{Entem:2003ft},
         the CD Bonn 2000 potential of Ref.~\cite{Machleidt:2000ge} and the potential developed 
         by Gross and Stadler in Ref.~\cite{Gross:2008ps}. Also shown are results
         from the Nijmegen partial wave analysis \cite{Stoks:1993tb}. 
         References to data can be found in  \cite{Stoks:1993tb}.}
\label{fig:NN}
\end{figure}

The resulting contributions at various orders to  the 2N, the 3N and the 4N forces
are depicted in Fig.~\ref{fig:power}.  Remarkably, by now the
2N, 3N and 4N force contributions have been worked out to N$^3$LO,
the last missing piece, namely the N$^3$LO corrections to the 3N forces, 
was only provided recently \cite{Ishikawa:2007zz,Bernard:2007sp,Bernard:2011zr}. Note, however, that the 3N forces might not
have fully converged at this order, and therefore a systematic study
of N$^4$LO contributions is underway by the Bochum group \cite{Krebs:2012yv,Krebs:2013kha}.
This EFT approach shares a few advantages over the very well
developed and precise semi-phenomenological approaches, just to mention
the consistent derivation of 2N, 3N and 4N forces as well as electroweak
current operators, the possibility to work out theoretical uncertainties 
and to improve the precision by going to higher orders and, of course, the direct
connection to the spontaneously and explicitly broken chiral symmetry of QCD.
There has been a large body of work on testing and developing these
forces in few-nucleon systems, for comprehensive reviews see 
\cite{Epelbaum:2008ga,Machleidt:2011zz}.  As an appetizer, I show in Fig.~\ref{fig:NN}
the description of two-nucleon scattering observables, namely the neutron-proton differential cross
section and the analyzing power at $E_{\rm lab} = 50$~MeV, in this type of approach compared to more
conventional and less systematic meson-exchange models.

\section{Ab initio solution of the nuclear many-body problem}
\label{sec:nls}

For systems up to four nucleons, one can calculate their properties
using the Faddeev-Yakubowsky machinery or using hyperspherical harmonics 
or other well developed methods. However, since we are interested in 
carbon and oxygen, we also have to consider  the nuclear many-body problem, 
which refers to nuclei with atomic number $A > 4$. The most modern tool 
to be used here are the so-called {\em nuclear lattice simulations}. They  combine the power of 
EFT to generate few-nucleon forces with computational methods to exactly solve the
non-relativistic $A$-body system, where in a nucleus $A$ counts the
number of neutrons plus protons. The basic ideas and definitions are 
spelled out in Ref.~\cite{Borasoy:2006qn} and for a detailed review
on lattice methods for non-relativistic systems, I refer to
Ref.~\cite{Lee:2008fa}. Here I give only a very short account 
of this method.  
The basic idea is is to discretize space-time and to introduce 
a smallest length (the lattice spacing) in the spatial directions 
and in the temporal direction, denoted $a$ and $a_t$, respectively.
The world is thus mapped onto a finite space-time volume  $L\times L\times L\times L_t$ 
in  integer multiples of $a$ and $a_t$, so $L = Na$ and $L_t=N_ta_t$, respectively.
 Typical values are $N=6$ and $N_t= 10\ldots 15$. A Wick rotation to Euclidean
space is naturally implied. Note that the finite  lattice spacing $a$ entails an ultra-violet
(UV) cutoff (a maximal momentum), $p_{\rm max} = \pi/a$. In typical
simulations of atomic nuclei, one has $a\simeq 2\,$fm and thus
 $p_{\rm max} \simeq 300\,$MeV. In contrast to lattice QCD, the continuum
limit $a\to 0$ is not taken. This formulation
allows to calculate the correlation function 
\begin{equation}
Z(t) = \langle \psi_A |  \exp(-t H) |\psi_A \rangle ~,
\end{equation}
where $t$ is the  Euclidean time, $H$ the nuclear Hamiltonian constructed along the
lines described in Sec.~\ref{sec:nucl} and $|\psi_A \rangle$ an $A$--nucleon state.
Using standard methods, one can derive any observable from the correlation
function, e.g. the ground-state energy is simply the infinite time limit
of the logarithmic derivative of $Z(t)$ with respect to the time. Similarly,
excited states can be generated by  starting with an ensemble of
standing waves, generating a correlation matrix 
$Z^{ji}(t) = \langle \psi_A^j |  \exp(-t H) |\psi_A^i \rangle$, which
upon  diagonalization generates
the ground and excited states - the larger the initial state basis, the
more excited states can be extracted. The initial states are standing waves, projected onto the
proper quantum numbers of spin and parity. From these standing waves, the general wave functions 
$\psi_j(\vec n\,)$ ($j=1,\ldots , A$) with well-defined momentum using all possible translations,
$L^{-3/2} \sum_{\vec m} \psi_j(\vec n + \vec m\,)\exp(i \vec{P}\cdot \vec{m})$, can be constructed. Thus,
the center-of-mass problem is taken care of. Another recently developed method
is based on more complicated initial position-space wave functions \cite{Epelbaum:2012qn}.
A proper choice for the $\psi_j$ allows one to prepare certain types of
initial states,  such as shell-model wave functions, which can be symbolically written as
(of course, proper antisymmetrization has to be performed)
\begin{equation}
\psi_j (\vec n\,) = \exp[-c {\vec n}^2]~,~~
\psi_j' (\vec n\,) = n_x \exp[-c {\vec n}^2]~,~~
\psi_j'' (\vec n\,) = n_y \exp[-c {\vec n}^2]~, \dots~,
\end{equation}
or, for later use, alpha-cluster wave functions,
\begin{equation}
\psi_j (\vec n\,) = \exp[-c (\vec n - \vec{m})^2]~,~~
\psi_j' (\vec n\,) = \exp[-c (\vec n -\vec{m}')^2]~,~~
\dots~,
\end{equation}
where $\vec n$, $\vec m$, $\ldots$ are triplets of integers that represent
a lattice site, and $n_x, n_y, \ldots$ the components of these vectors.
The possibility to construct all these different types of initial/final states
is a  reflection of the fact that in the underlying EFT all possible
configurations to distribute nucleons over all lattice sites are generated.
This includes in particular the configuration where four nucleons are
located at one lattice site, so there is no restriction like e.g. in 
a no-core-shell model approach, in which one encounters serious problems
with the phenomenon of clustering, that is so prominent in nuclear physics.
It is also important to note that the nuclear forces have an approximate
spin-isospin SU(4) symmetry (Wigner symmetry) \cite{Wigner:1936dx}
that is of fundamental importance in suppressing the malicious sign oscillations 
that plague any Monte Carlo simulation of strongly interacting Fermion systems 
at finite density. The relation of the Wigner symmetry to the nuclear EFT 
formulation has been worked out in Ref.~\cite{Mehen:1999qs} and its
consequences for lattice simulations are explored in Refs.~\cite{Chen:2004rq,Lee:2007eu}.

As one application of this method, I want to discuss the spectrum of
$^{12}$C and in particular the Hoyle state. This excited
state has been an enigma for nuclear structure theory since decades, even
the most successful Greens function MC methods based on realistic two- and
three-nucleon forces \cite{Pieper:2007ax} or the no-core-shell-model 
employing modern (renormalization group softened chiral) interactions
\cite{Navratil:2007we,Roth:2011ar} have not been able to describe this state.
The first {\em ab initio} calculation of the Hoyle state based on nuclear
lattice simulations was reported in Ref.~\cite{Epelbaum:2011md}. In the
meantime, the calculation of the spectrum  and the structure of $^{12}$C has been considerably
improved, using the aforementioned position-space initial and final state
wave functions \cite{Epelbaum:2012qn}. The predictions for the even-parity states in the $^{12}$C 
spectrum are collected in Tab.~\ref{tab:12C}. In all cases, the LO calculation
is within 10\% of the experimental number, and the three-nucleon forces at 
NNLO are essential to achieve agreement with experiment.  We remark, however, that the
so-called leading order subsumes various important higher order corrections, since
the LO four-nucleon contact interactions are smeared with a Gaussian-type function as 
discussed in Ref.~\cite{Borasoy:2006qn}.  The Hoyle state 
is clearly recovered and comes out at almost the same energy as the
$^4$He+$^8$Be threshold - thus allowing for the resonant enhancement
of carbon production that
was first considered by Hoyle half a century ago. Furthermore,
one finds a second $2^+$ excited state that has been much debated in the
literature. It agrees with the most recent determinations  \cite{Zimmerman:2013cxa}.
It is worth to stress that the method has been improved since the results shown
in Tab.~\ref{tab:12C} have been obtained. The ground state energy of $^{12}$C 
can now be calculated with an accuracy of about 200~keV  \cite{Lahde:2013uqa}.
 As already pointed out, the chiral nuclear EFT will also allow one to investigate
how the closeness of 
the Hoyle state to the $^4$He+$^8$Be thres\-hold depends on the fundamental
parameters like the light quark masses, thus allowing for a test of the
anthropic principle. For a first attempt within an alpha-cluster model,
see Ref.~\cite{Oberhummer:2000zj}.
\begin{table}
\begin{center}
\caption{The even-parity spectrum of $^{12}$C from nuclear lattice
simulations. The ground state is denoted as $O_1^+$ and the Hoyle
state as $O_2^+$. The NLO corrections include strong isospin breaking
as well as the Coulomb force. The NNLO corrections are generated by the
leading three-nucleon forces.  The theoretical errors include both Monte
Carlo statistical errors and uncertainties due to extrapolation at large
Euclidean time.
}
\begin{tabular}{|c|c|c|c|c|}
\hline
  & $0_1^+$ &  $2_1^+$ &  $0_2^+$ &  $2_2^+$ \\
\hline
LO    & $-96(2)$~MeV & $-94(2)$~MeV & $-88(2)$~MeV& $-84(2)$~MeV\\
NLO   & $-77(3)$~MeV & $-72(3)$~MeV & $-71(3)$~MeV& $-66(3)$~MeV \\
NNLO  & $-92(3)$~MeV & $-86(3)$~MeV & $-84(3)$~MeV& $-79(3)$~MeV\\
\hline
Exp.  & $-92.2$~MeV & $-87.7$~MeV & $-84.5$~MeV & $-82.2(1)$~MeV \cite{Zimmerman:2013cxa}\\
\hline
\end{tabular}
\label{tab:12C}
\end{center}
\end{table}

So far, nuclear lattice simulations have been performed at NNLO, which
includes the leading and dominant three-nucleon force topologies, see
Refs.~\cite{Epelbaum:2002vt,Epelbaum:2009zsa}. For nuclei up to carbon-12,
this is a good approximation due to the small cut-off $\Lambda = p_{\rm max} \simeq 300\,$MeV,
which is a much softer interaction than used in the description of continuum
NN scattering. Still, higher orders have eventually to be included to reduce the
theoretical uncertainties. Also, going to heavier nuclei one observes some 
overbinding with these NNLO forces \cite{Lahde:2013uqa} that grows with atomic number $A$.
This also requires the inclusion of higher order corrections to the two- and three-nucleon
forces. Work in this direction is under way.

\section{The nuclear force at varying quark mass}
\label{sec:forces}

\begin{figure}[t]
\begin{center}
 \includegraphics[width=0.75\textwidth]{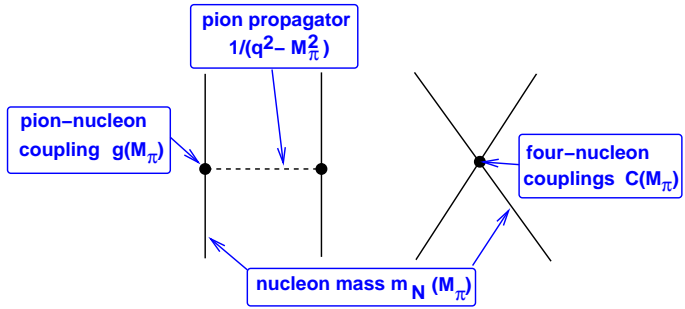}
\end{center}
\caption{Explicit and implicit pion (quark) mass dependence of the
leading order nucleon-nucleon (NN) potential. Solid (dashed) lines denote
nucleons (pions).}
\label{fig:mpi}
\end{figure}

\begin{figure}[t]
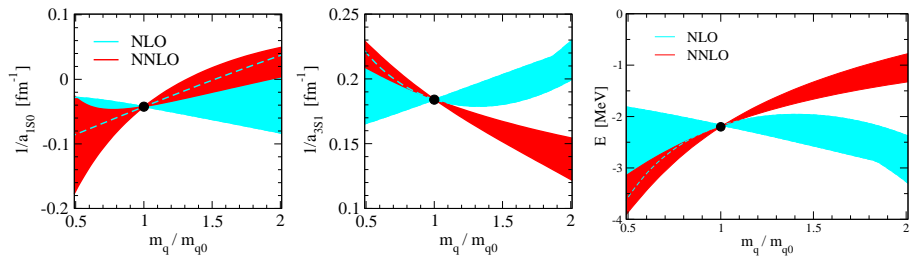

 \includegraphics[width=0.63\textwidth]{InvScatt_RunTPE.eps}~~
 \includegraphics[width=0.35\textwidth]{BE_RunTPE.eps}
\caption{Quark mass dependence of the inverse scattering length $1/a_{1S0}$ (left panel)
and $1/a_{3S1}$ (center panel) and the deuteron binding energy (right panel). Here, $m_{q0}$ denotes the
physical light quark mass.}
\label{fig:run}
\end{figure}

In the Weinberg approach to the nuclear forces, the quark mass
dependence of these forces can be worked out straightforwardly. To be precise, one encounters
{\it explicit} and {\it implicit} quark mass dependences. While the former
are generated through the  pion propagator, the latter stem from the quark mass
dependence  of the pion-nucleon coupling constant $\sim g_A/(2F_\pi)$, the nucleon mass, 
and the  4N couplings, respectively, see 
Fig.\ref{fig:mpi}.
Throughout, we use the Gell-Mann--Oakes--Renner relation,
\begin{equation}
 M_\pi^2 =  B(m_u+m_d) + {\cal O}((m_u+m_d)^2)~,
 \end{equation} 
 with $B$ a low-energy constant related to the scalar quark condensate. 
 In QCD, this relation is fulfilled to about 
 95\% \cite{Colangelo:2001sp},
so one can use the wording pion and quark mass dependence synonymously.
 For any observable ${\cal O}$ of a hadron $H$, we can define
its quark mass dependence in terms of the so-called $K$-factor, 
\begin{equation}
\frac{\delta {\cal O}_H}{\delta m_f}  \equiv K_H^f \,\frac{{\cal O}_H}{m_f}~, 
\end{equation} 
with $f=u,d$,
and $m_f$ the corresponding quark mass.
The pion mass dependence of pion and nucleon properties can be obtained from
lattice QCD combined with chiral perturbation theory as detailed in
Ref.~\cite{Berengut:2013nh}$\,$. The pertinent results are: 
\begin{equation}
K_{M_\pi}^q = 0.494^{+0.009}_{-0.013}~, \quad 
K_{F_\pi}^q = 0.048\pm 0.012~, \quad
K_{m_N}^q = 0.048^{+0.002}_{-0.006}~, 
\end{equation}
where $q$ denotes the average light quark mass. To a good approximation,
$K^q_{g_A} \simeq 0$. 
For the quark mass dependence of the short-distance terms, 
one has to resort to modeling using resonance saturation~\cite{Epelbaum:2001fm}. This induces
a sizeable uncertainty that might be overcome by lattice QCD simulations in the
future. For the NN scattering lengths and the deuteron binding energy (BE), this leads to 
\begin{equation}
K^q_{1S0} = 2.3^{+1.9}_{-1.8}~, \quad
K^q_{3S1} = 0.32^{+0.17}_{-0.18}~, \quad
K^q_{\rm BE(deut)} = -0.86^{+0.45}_{-0.50}~, 
\end{equation}
extending and  improving earlier work based on EFTs and 
models~\cite{Muther:1987sr,Beane:2002vs,Epelbaum:2002gb,Flambaum:2007mj,Soto:2011tb}.
The running of the NN scattering lengths and the deuteron BE with the light quark 
mass is shown in Fig.~\ref{fig:run}. Note, however, that there are recent lattice QCD
simulations at large pion masses of about 500 and 800~MeV that seem to indicate a
decrease of the deuteron BE with pion mass \cite{Yamazaki:2009ua,Beane:2012vq}.
How solid extrapolations from such large values down to the physical pion mass 
are, remains however questionable.  
In addition to shifts in $m_q^{}$, we shall also consider the
effects of shifts in $\alpha_{\rm EM}^{}$. The
treatment of the Coulomb interaction in the nuclear lattice EFT framework is
described in detail in Ref.~\cite{Epelbaum:2010xt}.
\section{Constraints from Big Bang Nucleosynthesis}
\label{sec:bbn}

Using the results from the previous section, one can now analyze what
constraints the element abundances in BBN on possible quark mass variations imply.
At the beginning, we keep the electromagnetic fine structure constant fixed and
work in the isospin limit $m_u=m_d=m_q$. In BBN, elements up to $^7$Li and
$^7$Be are produced, but in what follows we consider only the variation
of the NN scattering lengths, the deuteron BE and  we also need the variation of the
BEs of $^3$He and $^4$He with the pion mass. Following Bedaque, Luu and Platter
 (for short BLP)~\cite{Bedaque:2010hr}
 these can be obtained by convoluting the 2N $K$-factors with the variation of the
3- and 4-particle BEs with respect to the singlet and triplet NN scattering 
lengths. This gives
\begin{equation}
 K_{^3{\rm He}}^q = -0.94\pm 0.75~, \quad K_{^3{\rm He}}^q  = -0.55\pm 0.42~,
\end{equation}
for details I refer to Ref.~\cite{Berengut:2013nh}. These  values are consistent 
with a direct
calculation using nuclear lattice simulations, $K_{^3{\rm He}}^q = -0.19\pm
0.25$ and $K_{^3{\rm He}}^q = -0.16\pm 0.26$~\cite{Lahde}. With this input,
we can calculate the BBN response matrix of the primordial abundances $Y_a$ at 
fixed baryon-to-photon ratio, 
\begin{equation}
\frac{\delta Y_a}{\delta m_q} = \sum_{X_i} \frac{\delta \ln Y_a}{\delta \ln X_i}\, K_{X_i}^q~, 
\end{equation}
with $X_i$ the relevant BEs for $^2$H,
$^3$H, $^3$He, $^4$He, $^6$Li, $^7$Li and  $^7$Be and the singlet NN
scattering length, using the updated Kawano code (for details,
see Ref.~\cite{Berengut:2009js}). Comparing the calculated with the observed
abundances, one finds that the most stringent limits arise from the
deuteron abundance [deut/H] and the $^4$He abundance normalized to the one
of protons, $^4$He($Y_p$), as most neutrons end up in the alpha
nucleus. Combining these leads to the constraint {$\delta m_q/m_q = (2\pm 4)\%$}.
These values are consistent with earlier determinations based on models of the
nuclear forces.
In contrast to these earlier determinations, we provide reliable error
estimates due to the underlying EFT. However, as pointed out by BLP, one
can obtain an even stronger bound due to the neutron lifetime, which 
strongly affects  $^4$He($Y_p$).  To properly address this issue, one has
of course to include strong isospin violation, as the neutron-proton mass
difference receives a  2~MeV contribution from the light quark mass difference
and about $-0.7$~MeV from the electromagnetic interactions. Re-evaluating this constraint 
under the model-independent assumption that {\em all} quark and lepton masses
vary with the Higgs vacuum expectation value (VEV) $v$, leads to 
\begin{equation}\label{eq:bbn}
\left|\frac{\delta v}{v}\right| = \left|\frac{\delta m_q}{m_q}\right| \leq 0.9\%~. 
\end{equation}
This is similar to what has been found by BLP, however, they assumed that when $m_q$
changes, $m_u/m_d$ and all other Standard Model parameters stay constant. Such a
scenario is hard to reconcile with the Higgs mechanism that gives mass to all fundamental
particles $\sim v$ - it would require some very intricate fine-tuning of Yukawa couplings.
Constraints on the variations of the Higgs VEV from nuclear binding have also been considered
in Ref.~\cite{Damour:2007uv}. Also, very recently bounds on quark mass and $\alpha_{\rm EM}$ variations
from an ab initio calculation of the neutron-proton mass difference have been reported \cite{Borsanyi:2014jba}.

\section{The fate of carbon-based life as a function of the fundamental parameters of the 
Standard Model}
\label{sec:fate}

\begin{figure}
\begin{center}
\includegraphics[width=0.95\textwidth]{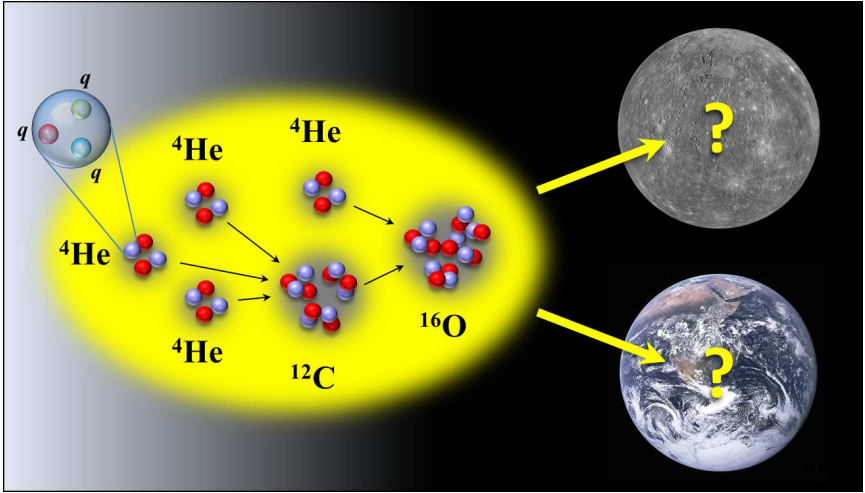}
\end{center}
\caption{Graphical representation of the question of how fine-tuned 
life on Earth is under variations of the average light quark mass and $\alpha_{\rm EM}$. 
Figure courtesy of Dean Lee.}
\label{fig:fate}
\end{figure}

I now turn to the central topic of this review, namely how fine-tuned is the
production of carbon and oxygen with respect to changes in the fundamental
parameters of QCD+QED? Or, stated differently, how much can we detune these
parameters from their physical values to still have an habitable Earth as shown
in Fig.~\ref{fig:fate}. To be more precise, we must specify which parameters
we can vary. In QCD, the strong coupling constant is tied to the nucleon mass
through dimensional transmutation. However, the light quark mass (here, only
the strong isospin limit is relevant) is an external parameter. Naively, one
could argue that due to the small contribution of the quark masses to the
proton and the neutron mass, one could allow for sizeable variations. However,
the relevant scale to be compared to here is the average binding energy per
nucleon, $E/A \leq 8\,$MeV (which is much smaller than the nucleon mass). 
As noted before, the Coulomb repulsion between protons is an important
ingredient in nuclear binding, therefore we must also consider changes in
$\alpha_{\rm EM}$. Therefore, in the following we will consider variations 
in the light quark mass $m_q$ at fixed fine structure constant $\alpha_{\rm EM}$
and also changes in $\alpha_{\rm EM}$ at fixed $m_q$.
The tool to do this are nuclear lattice simulations, which
allowed e.g. for the first {\em ab initio} calculation of the Hoyle 
state~\cite{Epelbaum:2011md}.

\begin{figure}[t]
\begin{center}
\includegraphics[width=0.48\textwidth]{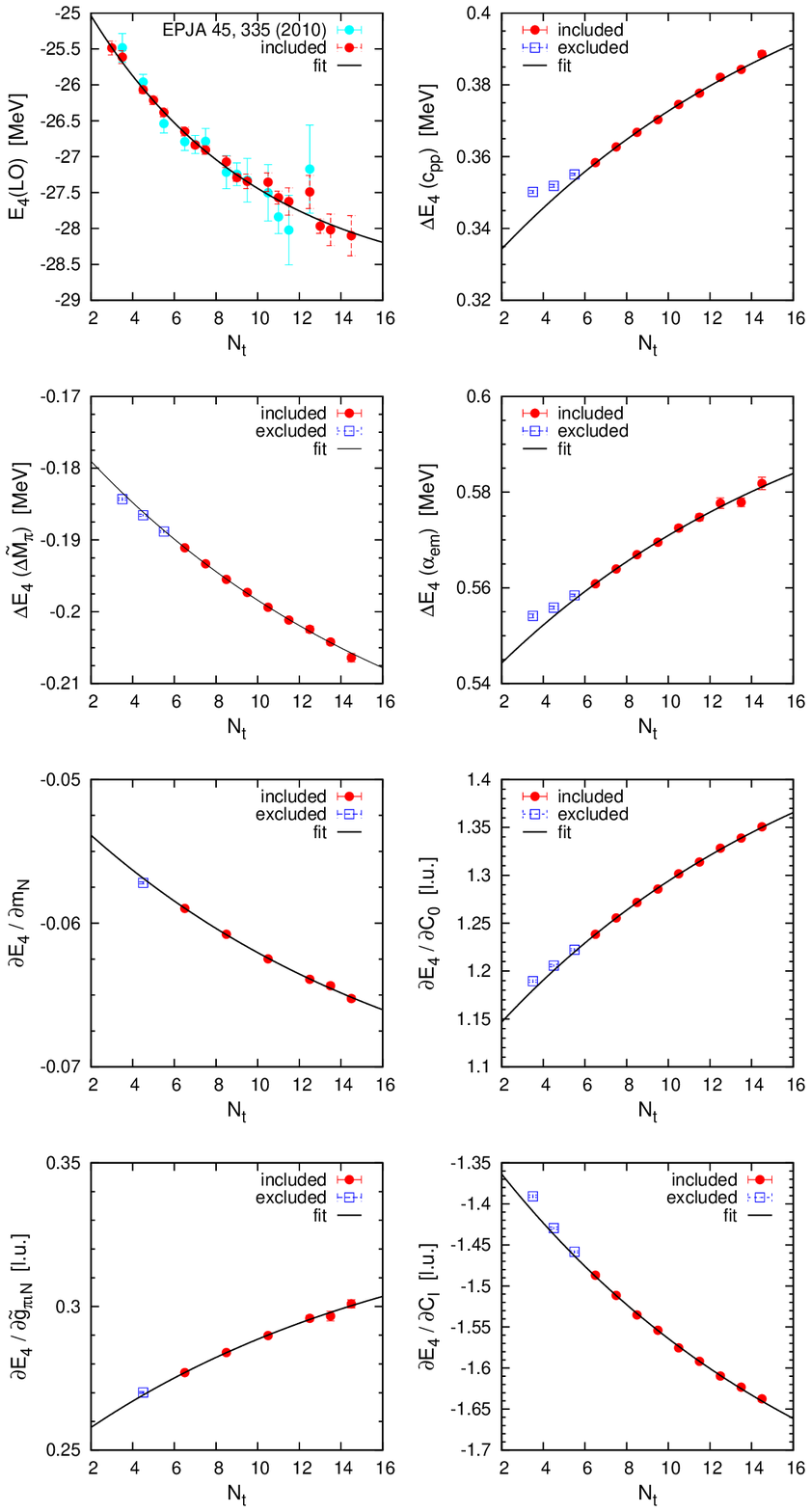}~~
\includegraphics[width=0.48\textwidth]{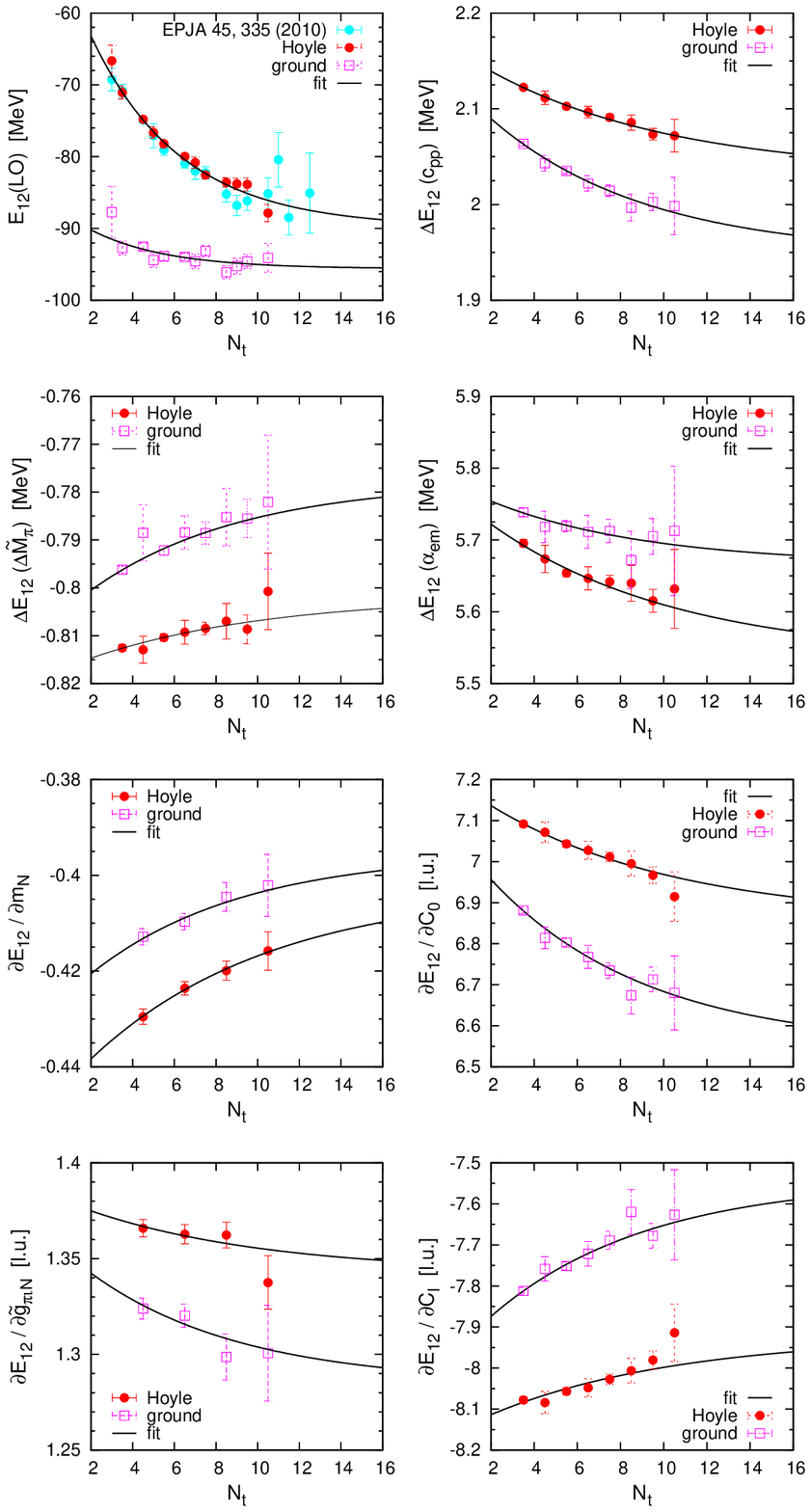}
\end{center}
\caption{AFQMC calculation of $^4$He (left panels)  and $^{12}$C (right panels), as a 
function of Euclidean time steps $N_t^{}$. For definition of the various energies and
derivatives shown, see Ref.~\cite{Epelbaum:2013wla}.
The results of Ref.~\cite{Epelbaum:2010xt} for $E_{12}^\star$(LO) are included 
to highlight the improved statistics, and as a consistency check.}
\label{fig:AFQMC}
\end{figure}

\begin{figure}[t]
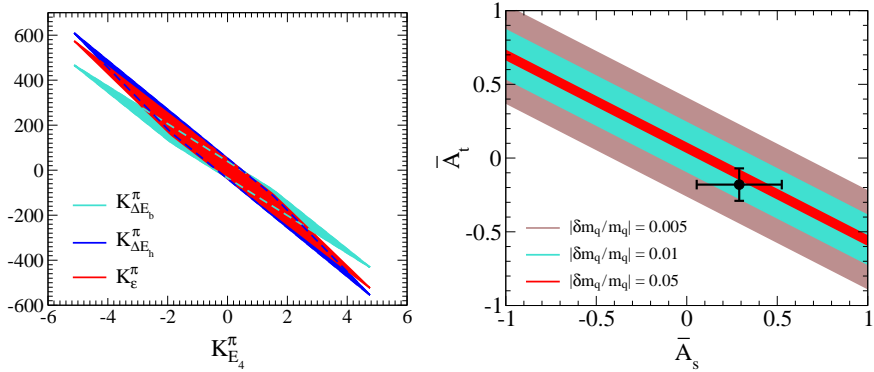

\begin{center}
 \includegraphics[width=0.45\textwidth]{corelDeltaBEv5.eps}~~
 \includegraphics[width=0.485\textwidth]{hoyle_band_new.eps}
\end{center}
\caption{Left panel: Sensitivities of $\Delta E_h^{}$, $\Delta E_b^{}$ and $\varepsilon$ 
to changes in $M_\pi^{}$, as a function of $K_{E_4^{}}^\pi$ under independent 
variation of $\bar A_s^{}$ and $\bar A_t^{}$ over the range $\{-1 \ldots 1\}$.   
The bands correspond to $\Delta E_b^{}$, 
$\varepsilon$ and  $\Delta E_h^{}$ in clockwise order.
Right panel: 
``Survivability bands'' for carbon-oxygen based life from
  Eq.~(\ref{final_res}), due to  $0.5\%$ (broad outer band), $1\%$ (medium
  band) and $5\%$ (narrow inner band) changes in $m_q^{}$ in terms of the
  input parameters $\bar A_s^{}$ and $\bar A_t^{}$. The most up-to-date 
  N$^2$LO analysis of $\bar A_s^{}$ and $\bar A_t^{}$ from
  Ref.~\cite{Berengut:2009js} is given by the data point with 
  horizontal and vertical error bars.}
\label{fig:band}
\end{figure}

 Let us consider first QCD, i.e. variations in the light quark mass  at fixed $\alpha_{\rm EM}$
(for details, see Refs.~\cite{Epelbaum:2012iu,Epelbaum:2013wla}). We want to calculate
the variations of the pertinent energy differences  in the
triple-alpha process $\delta \Delta E/ \delta M_\pi$, which according to
Fig.~\ref{fig:mpi} boils down to (we consider small variations around
the physical value of the pion mass $M_\pi^\mathrm{ph}$):
\begin{eqnarray}
\left. \frac{\partial E_i^{}}{\partial M_\pi^{}} \right|_{M_\pi^\mathrm{ph}} &= &
\left. \frac{\partial E_i^{}}{\partial \tilde M_\pi^{}} \right|_{M_\pi^\mathrm{ph}}
+ x_1^{} \left. \frac{\partial E_i^{}}{\partial m_N^{}} \right|_{m_N^\mathrm{ph}}
+ x_2^{} \left. \frac{\partial E_i^{}}{\partial g_{\pi N}^{}} \right|_{g_{\pi N}^\mathrm{ph}} 
\nonumber \\
& +& x_3^{} \left. \frac{\partial E_i^{}}{\partial C_0^{}}
\right|_{C_0^\mathrm{ph}}
+ x_4^{} \left. \frac{\partial E_i^{}}{\partial C_I^{}} \right|_{C_I^\mathrm{ph}},
\label{Eeq2}
\end{eqnarray}
with the definitions
\begin{equation}
x_1^{} \equiv \left. \frac{\partial m_N^{}}{\partial M_\pi^{}} 
\right|_{M_\pi^\mathrm{ph}}, ~~
x_2^{} \left. \equiv \frac{\partial g_{\pi N}^{}}{\partial M_\pi^{}}
\right|_{M_\pi^\mathrm{ph}}~
x_3^{} \equiv \left. \frac{\partial C_0^{}}{\partial M_\pi^{}}
\right|_{M_\pi^\mathrm{ph}}, ~~
x_4^{} \equiv \left. \frac{\partial C_I^{}}{\partial M_\pi^{}} \right|_{M_\pi^\mathrm{ph}},
\label{xy}
\end{equation}
with $\tilde{M_\pi}$ the pion mass appearing in the pion-exchange potential. 
The various derivatives in Eq.~(\ref{Eeq2}) can be obtained precisely using
Auxiliary Field Quantum Monte Carlo (AFQMC) techniques, as examples we show the
various contributions to the energy and the
various derivatives for $^4$He and $^{12}$C in Fig.~\ref{fig:AFQMC}.
The $x_i$ ($i = 1,2,3,4$) are related to
the pion and nucleon as well as the two-nucleon  $K$-factors determined in Sec.~\ref{sec:forces}. 
As described in detail in Ref.~\cite{Epelbaum:2013wla}, the current knowledge of the quark mass
dependence of the nucleon mass, the pion decay constant and the pion-nucleon coupling constant
leads to $x_1 = 0.57\ldots 0.97$ and $x_2 = -0.056 \ldots 0.008$ (in lattice units).
The scheme-dependent quantities $x_{3,4}$ can be traded for the
pion-mass dependence of the inverse singlet $a_s$ and triplet $a_t$ scattering lengths,
\begin{equation}
\bar A_{s}^{} \equiv \frac{\partial a_{s}^{-1}}{\partial M_\pi}\biggr|_{M_\pi^{\rm ph}}~,  \quad
\bar A_{t}^{} \equiv \frac{\partial a_{t}^{-1}}{\partial M_\pi}\biggr|_{M_\pi^{\rm ph}}~.
 \end{equation}   
We can then
express all energy differences appearing in the triple-alpha process 
\begin{equation}
\Delta E_b^{} \equiv E_8^{} - 2 E_4^{}~, \quad
\Delta E_h^{} \equiv E_{12}^\star - E_8^{} - E_4^{}~,\quad
\varepsilon \equiv E_{12}^\star - 3E_4^{}~,
\end{equation}
where $E_4^{}$ and $E_8^{}$ denote the energies of 
the ground states of $^{4}$He and $^{8}$Be, respectively, and $E_{12}^\star$ denotes the
energy of the Hoyle state, as functions
of $\bar A_{s}$ and $\bar A_{t}$. One finds that all these energy differences
are correlated, i.e. the various fine-tunings in the triple-alpha process are not independent
of each others, see the left panel of Fig.~\ref{fig:band}. Further, one finds a 
strong dependence on the variations of the
$^4$He BE, which is strongly suggestive of the $\alpha$-cluster structure of
the $^8$Be, $^{12}$C and Hoyle states.  Such correlations related to the production 
of carbon have indeed been speculated upon earlier~\cite{Livio,WeinbergFacing},
but only with the techniques displayed here one could finally derive them from first principles.

Consider now the reaction rate of the triple-alpha process as given by
\begin{equation}
\label{rate}
r_{3 \alpha}^{} = 3^{\frac{3}{2}} N_\alpha^3 
\left( \frac{2 \pi \hbar^2}{|E_4^{}| k_{\rm B}^{} T} \right)^3 
\frac{\Gamma_\gamma^{}}{\hbar} \, \exp \left( -\frac{\varepsilon}{k_{\rm B}^{} T} \right)~,  
\end{equation}
with $N_\alpha$ the $\alpha$-particle number density in the stellar plasma with temperature $T$, 
$\Gamma_\gamma = 3.7(5)\,{\rm meV}$ the radiative width of the Hoyle state and 
$k_B$ is Boltzmann's constant.
The stellar modeling calculations of Refs.~\cite{Oberhummer:2000mn,Oberhummer-astro} suggest 
that sufficient abundances of both carbon and oxygen can be maintained within an envelope of 
$\pm 100$~keV around the empirical value of $\varepsilon = 379.47(18)$~keV. 
This condition can be turned into a constraint on shifts in $m_q^{}$ that reads
(for more details, see Ref.~\cite{Epelbaum:2013wla})
\begin{equation}
\left| \Big[ 0.572(19) \, \bar A_s^{} + 0.933(15) \, \bar A_t^{} 
- 0.064(6)  \Big]  
\left(\frac{\delta m_q^{}}{m_q^{}} \right) \right|  < 0.15\%~.
\label{final_res}
\end{equation}
The resulting constraints on the values of $\bar
A_s^{}$ and $\bar A_t^{}$ compatible
with the condition $| \delta \varepsilon | < 100$~keV are visualized in the
right panel of Fig.~\ref{fig:band}.  The various shaded bands in this figure cover the 
values of $\bar A_s^{}$ and $\bar A_t^{}$ consistent
with carbon-oxygen based life, when $m_q^{}$ is varied by $0.5$\%, $1$\% and $5$\%.
Given the current theoretical 
uncertainty in $\bar A_s^{}$ and $\bar A_t^{}$, our results remain compatible 
with a vanishing $\partial \varepsilon / \partial M_\pi^{}$, in other words
with a complete lack of fine-tuning. Interestingly, Fig.~\ref{fig:band} (right panel) 
also  indicates that the triple-alpha process is unlikely to be fine-tuned
to a higher degree than $\simeq 0.8$\% under variation of $m_q^{}$. 
The central values of $\bar A_s^{}$ and $\bar A_t^{}$ from
Ref.~\cite{Berengut:2009js} suggest that variations in the light quark masses 
of up to $2 - 3$\% are unlikely to be catastrophic to the formation of 
life-essential carbon and oxygen. A similar calculation of the tolerance for 
shifts in the fine-structure constant $\alpha_{\rm EM}^{}$ proceeds as follows.
For small variations $| \delta \alpha_{\rm em}^{}/\alpha_{\rm em}^{} | \ll 1$ at the fixed 
physical value of $m_q$,  the resulting change in $\varepsilon$ can be expressed as
\begin{equation}
\delta (\varepsilon ) \approx 
\left. \frac{\partial \varepsilon}{\partial \alpha_{\rm em}^{}} \right |_{\alpha_{\rm em}^\mathrm{ph}} 
\delta \alpha_{\rm em}^{} 
=  
Q_{\rm em}^{}(\varepsilon) 
\left(\frac{\delta \alpha_{\rm em}^{}}{\alpha_{\rm em}^{}}\right)~,
\end{equation}
where $Q_{\rm em} (\varepsilon) = 3.99(9)$~MeV receives contributions from
the long-range Cou\-lomb force and a $pp$ contact term (for details, we refer to
Ref.~ \cite{Epelbaum:2013wla}). Recalling further that $K_{M_\pi^{}}^q =
0.494^{+0.009}_{-0.013}$~\cite{Berengut:2013nh},  
the  condition $| \delta (\varepsilon )| < 100$~keV leads to the predicted tolerance
$| \delta \alpha_{\rm em}^{}/\alpha_{\rm em}^{} | \simeq 2.5 \%$ of 
carbon-oxygen based life to shifts in $\alpha_{\rm em}^{}$.
This result is compatible with the $\simeq 4\%$ bound reported 
in Ref.~\cite{Oberhummer:2000mn}.   

\section{A short discussion of the anthropic principle}
\label{sec:ant}

Let us pause and discuss the findings obtained in the previous sections. First,
it is important to stress that we only consider deformations of the Standard Model
that can be expressed through variations of  the light quark mass and the 
electromagnetic fine structure constant. 
One could imagine a completely different approach to the strong and electroweak
interactions that might also lead to carbon-oxygen based life given the proper
cosmological conditions. While that is certainly possible, it goes beyond the type
of tests we are after. Thus, we will not further consider such possibilities but rather
discuss our more modest approach. 

Consider
first the element generation in the Big Bang. From the observed element abundances and
the fact that the free neutron decays in about 882~s and the surviving neutrons are
mostly captured in $^4$He, one finds a stringent bound on the light quark mass variations
as given in Eq.~(\ref{eq:bbn}), under the reasonable assumption that the masses of all
quarks and leptons appearing in neutron $\beta$-decay scale with the Higgs VEV.
Thus, BBN sets indeed very tight limits on the variations of the light quark mass.
Such extreme fine-tuning supports the anthropic view of our Universe.

The situation concerning the fine-tuning in the triple-alpha process is somewhat less
clear. As noted already in Refs.~\cite{Livio,WeinbergFacing}, the allowed variations in $\varepsilon$
(remember that the size of $\varepsilon$ defines the resonance condition for carbon production)
are not that small, as $|\delta \varepsilon/\varepsilon | \simeq 25\%$ still allows
for carbon-oxygen based life. So one might argue that the anthropic principle is
indeed {\it not} needed to explain the fine-tunings in the triple-alpha process.
However, as we just showed, this translates into allowed quark mass variations
of $2-3\%$ and modifications of the fine-structure constant of about 2.5\%.
The  fine-tuning in the fundamental parameters is thus much more severe than the
one in the energy difference $\varepsilon$. Therefore, beyond such relatively small changes 
in the fundamental para\-meters, the anthropic principle indeed appears necessary to 
explain the observed abundances of $^{12}$C and $^{16}$O. Of course, to sharpen 
these statements, on must be able to reduce the uncertainty in the determination of
the quark mass dependence of the LO four-nucleon contact operators given by the
quantities $\bar A_s^{}$ and $\bar A_t^{}$. It is hoped that lattice QCD simulations of the
two-nucleon system will be able to reduce the sizable uncertainty in these parameters.

\section{Summary and outlook}
\label{sec:sum}

In this short review, I have summarized recent developments in our understanding of
the fine-tuning in the generation of the life-essential elements as well as
the light elements generated in BBN. As shown, the allowed parameter
variations in QCD+QED are small, giving some credit to the anthropic
principle. To sharpen these conclusions, future work is required. On one side,
lattice QCD at sufficiently small quark masses will eventually be able to
give tighter constraints on the parameters $\bar A_{s,t}$ and on the other
side, nuclear lattice simu\-lations have to be made more precise to further reduce the
theoretical error in the binding and excitation energies and to provide {\it
ab initio} calculations of nuclear reactions, for first steps, 
see Refs.~\cite{Rupak:2013aue,Pine:2013zja}$\,$. Finally, we remark that we 
have considered here QCD with a vanishing $\theta$-angle. For a recent study 
on variations of $\theta$ on the deuteron BE and the triple-alpha process, see
Ref.~\cite{Ubaldi:2008nf}.


\subsection*{Acknowledgments}

I am grateful to Steve Weinberg, whose query on the resonance condition triggered part
of the work done here.
I would like to thank my NLEFT collaborators Evgeny Epelbaum, Hermann Krebs, 
Timo L\"ahde, Dean Lee and also Gautam Rupak for a most enjoyable collaboration.  
Some part of this work was done in collaboration with Julian Berengut, Victor Flambaum,
Christoph Hanhart, Jenifer Nebreda and Jose Ramon Pel\'aez.
I would also like to thank  Zhizhong Xing for giving me the opportunity to write
this review.  I am grateful to Evgeny Epelbaum, Dean Lee and Qiang Zhao for a careful
reading of the manuscript.

\end{document}